\newif\ifabstract
\abstracttrue
 \abstractfalse %Uncommenting this line gives the full version
\newif\iffull
\ifabstract \fullfalse \else \fulltrue \fi
\documentclass[11pt]{article}
\usepackage{amsfonts}
\usepackage{amssymb}
\usepackage{amstext}
\usepackage{amsmath}
\usepackage{xspace}
\usepackage{theorem}
\usepackage{graphicx}
\usepackage{url}
\usepackage{graphics}
\usepackage{colordvi}
\usepackage{colordvi}
\usepackage{subfigure}

\advance \oddsidemargin by -0.8in
\newcommand{\myparskip}{3pt}
\parskip \myparskip

        {\hspace*{\fill}$\Box$\par\vspace{4mm}}

\newcommand{\be}{\begin{enumerate}}
\newcommand{\ee}{\end{enumerate}}
\newcommand{\bd}{\begin{description}}
\newcommand{\ed}{\end{description}}
\newcommand{\bi}{\begin{itemize}}
\newcommand{\ei}{\end{itemize}}

\def\stopproof{\square}
\def\square{\vbox{\hrule height.2pt\hbox{\vrule width.2pt height5pt \kern5pt
\vrule width.2pt} \hrule height.2pt}}

\renewcommand{\phi}{\varphi}

\mathchardef\hyphen="2D

\newcommand\blfootnote[1]{%
  \begingroup
  \renewcommand\thefootnote{}\footnote{#1}%
  \addtocounter{footnote}{-1}%
  \endgroup
}

\usepackage{enumitem}
\usepackage{multirow}
\usepackage[ruled,vlined]{algorithm2e}

\begin{document}

\title{Analysis of N-of-1 trials using Bayesian distributed lag model with autocorrelated errors}
\author{Ziwei Liao$^1$, Min Qian$^1$, Ian M. Kronish$^2$, Ying Kuen Cheung$^1$}
\date{}

\begin{titlepage}
\maketitle

\thispagestyle{empty}

\begin{abstract}
\normalsize
An N-of-1 trial is a multi-period crossover trial performed in a single individual, with a primary goal to estimate treatment effect on the individual instead of population-level mean responses. As in a conventional crossover trial, it is critical to understand carryover effects of the treatment in an N-of-1 trial, especially when no washout periods between treatment periods are instituted to reduce trial duration. To deal with this issue in situations where high volume of measurements is made during the study, we introduce a novel Bayesian distributed lag model that facilitates the estimation of carryover effects, while accounting for temporal correlations using an autoregressive model. Specifically, we propose a prior variance-covariance structure on the lag coefficients to address collinearity caused by the fact that treatment exposures are typically identical on successive days. A connection between the proposed Bayesian model and penalized regression is noted. Simulation results demonstrate that the proposed model substantially reduces the root mean squared error in the estimation of carryover effects and immediate effects when compared to other existing methods, while being comparable in the estimation of the total effects. We also apply the proposed method to assess the extent of carryover effects of light therapies in relieving depressive symptoms in cancer survivors.
\end{abstract}

\vspace{5mm}
\noindent \textit{Keywords:} \newline 
Bayesian distributed lag model, Carryover effects, N-of-1 trials, Personalized treatment,  Regression with autocorrelated errors,  Time series
\noindent

\blfootnote{$^1$Department of Biostatistics, Columbia University, New York, NY, USA}
\blfootnote{$^2$Center for Behavioral Cardiovascular Health, Columbia University, New York, NY, USA}
\blfootnote{\\ \textbf{Corresponding Author:}} 
\blfootnote{Ying Kuen Cheung, } 
\blfootnote{Department of Biostatistics, Mailman School of Public Health, Columbia University, New York, NY, USA.}
\blfootnote{Email: yc632@columbia.edu}

\end{titlepage}

\setcounter{secnumdepth}{2}

\section{Introduction}
\label{s:intro}

\begin{figure}
\begin{center}
\includegraphics[width=1.0\linewidth]{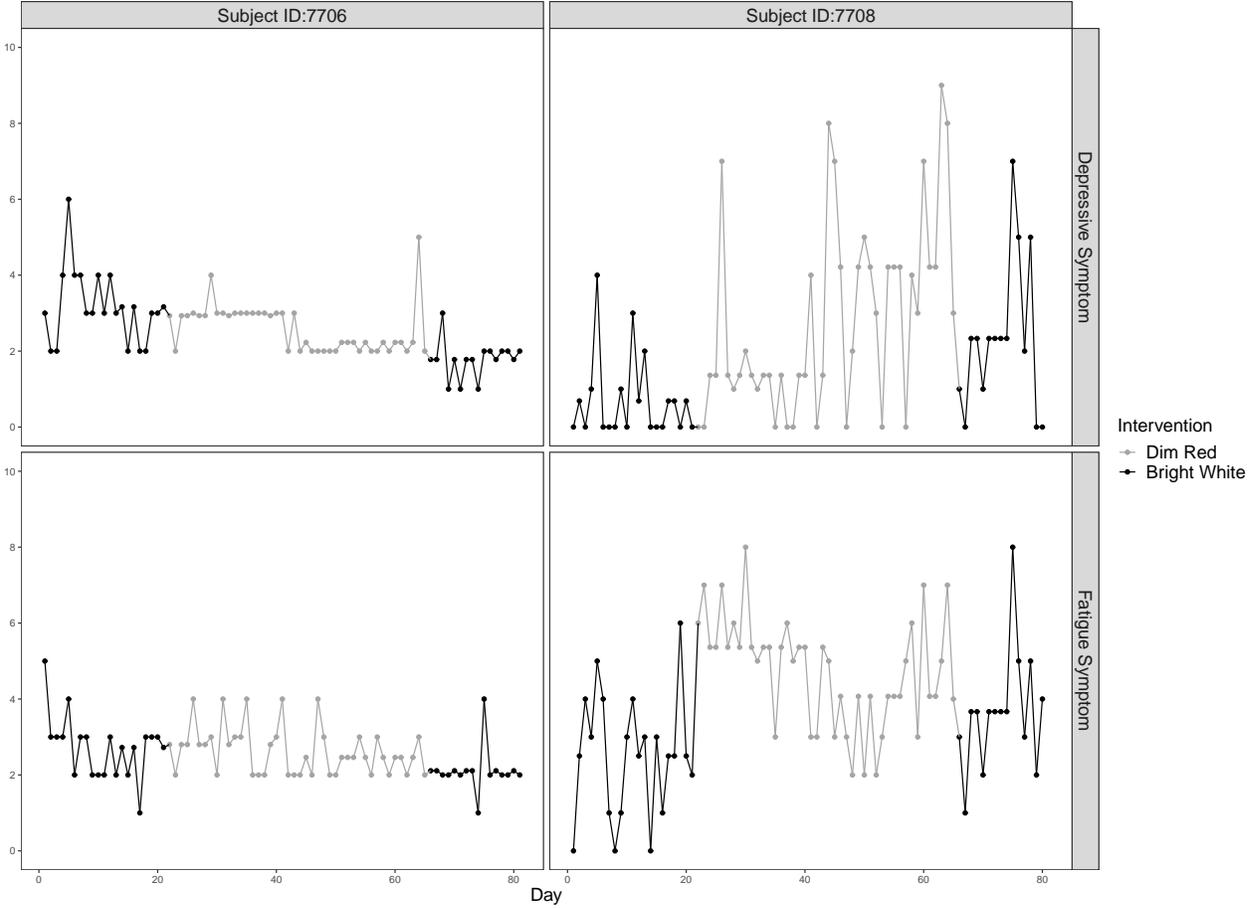}
\end{center}
\caption{Daily assessments of two patients id 7706 and 7708. Black line represents bright white light intervention, and grey line dim red light.}
\label{fig:timeSeries}
\end{figure}

N-of-1 trials are multi-period crossover studies that compare two or more interventions in single individuals, and are suitable for evaluating personalized treatment effects in those with chronic conditions where the outcome is relatively stable. \cite{kravitz2014design}
Advances in mobile and sensor technology \cite{topol2010transforming} and better understanding of patient preferences \cite{cheung2020personal} have improved the implementation of N-of-1 trials. However, their uptake remains very small in clinical practice. In particular, the duration of N-of-1 trials remains a key barrier. To reduce the duration needed to conduct an N-of-1 trial and to reduce the burden of participation, it is often necessary to preclude scheduling washout periods between treatments. When physical washout periods are not feasible, it is critical to have provisions for dealing with carryover effects analytically. 
To motivate our work, consider an N-of-1 trial series that compare bright white light (10,000 lux) and dim red light (50 lux) in cancer patients with depressive symptoms, where light therapy was delivered by portable light boxes with instructions. \cite{kronish2020clinical}
Briefly, each individual would use one of two light boxes for 30 minutes each morning over a 12 weeks. Along with the light boxes, a smartphone application would be used to give treatment reminders and to assess daily depressive symptoms and fatigue level over the entire 12-week period.  While theory suggests bright white light may reduce cancer-related depression and fatigue, its effects may vary from individual to individual. \cite{johnson2018bright}
Thus, the primary analytical goal in light therapy study is to identify for each individual whether bright white light is superior in terms of symptom control and make light therapy suggestion for their further clinical treatment. Figure~\ref{fig:timeSeries} shows the daily assessments of two patients during the study course.

In a systematic review of 108 N-of-1 trial series published between 1985 and 2010, Gabler et al. reported on the analytic methods used to compare the effectiveness of two or more treatments being studied in an N-of-1 trial, including graphical comparison, hypothesis tests (e.g., t-test, nonparametric tests), and regression models.\cite{gabler2011n}
While there is no single agreed upon analysis method, these methods ignore two key features of experimental N-of-1 data. 
First, most methods do not account for temporal dependence (i.e., autocorrelation) between assessments.
Second, the methods do not capture the carryover effects of an intervention.
The second data feature, which motivates this article, can be partly addressed by using a distributed lagged model (DLM), which is widely used in economics, \cite{koyck1954distributed, almon1965distributed} advertising, \cite{bass1972testing} and environmental health studies. \cite{welty2009bayesian, zanobetti2000generalized}
A DLM postulates that the current value of the outcome variable depends on the previous values (lags) of an exposure as well as the current exposure value, thus allowing the total exposure effect to be distributed over a time period and facilitating explicit modeling of carryover effects.
A potential challenge in fitting a DLM is collinearity of the exposure lags. The N-of-1 trial design will further aggravate the problem: as illustrated in Figure~\ref{fig:timeSeries}, the exposure (light box) often remains the same as in the previous day in order to avoid switching intervention too frequently during a trial.
A strategy to handle collinearity in DLM is by putting parametric constraints on the lag coefficients such as geometric lags, \cite{koyck1954distributed} or polynomial lags. \cite{almon1965distributed} Alternatively, one may consider putting informative prior on the coefficients in a Bayesian framework. \cite{welty2009bayesian}

In this article, we adopt the Bayesian framework and propose a Bayesian distributed lag model with autocorrelated errors (BDLM-AR) as an extension of DLMs for N-of-1 trial data. The model is novel in several ways. First, we propose a prior distribution that constrains the lag coefficients with shrinkage factors that increase over time. Second, we impose a fused ridge-type penalty to  address collinearity, which may be viewed as a variant of the fused lasso method. \cite{tibshirani2008spatial} Third, while current DLM methods assume independent error terms, we incorporate temporal correlations using an autoregressive error model. 
We will introduce the proposed BDLM-AR with details in Section \ref{s:model}, and describe the posterior computations in Section  \ref{s:posterior}. The performance of BDLM-AR will be evaluated and compared with other methods by simulation studies presented in Section \ref{s:simulation}. We will apply the proposed method to the light therapy data in Section \ref{s:application}, and will conclude this article with a discussion in Section \ref{s:discuss}.

\section{Bayesian Distributed Lag Model with Autocorrelated Error}
\label{s:model}
\subsection{Proposed Model}
Suppose we observe data from a patient on $n$ consecutive days. On day $t=1,\ldots, n$, let $X_t$ and $Y_t$ denote the binary treatment indicator and the outcome of interest, respectively.
We consider a distributed lag autoregressive model for $Y$, described as follows:
\begin{equation}
\label{eqn:1}
    Y_t= \mu + \sum^L_{l=0}\beta_l X_{t-l}+\epsilon_t
\end{equation}
for $t=p+1,...,n$, where the error term $\epsilon_t$ follows an autoregressive process,
\begin{equation}
\label{eqn:2}
    \epsilon_t=\phi_1\epsilon_{t-1}+\phi_2\epsilon_{t-2}+...+\phi_p\epsilon_{t-p}+w_t
\end{equation}
$w_t$ is a white Gaussian noise with mean zero and unknown variance $\sigma^2>0$, and $L$ and $p$ are pre-specified. Note that for $t<L$, the maximum lag effect is of order $t-1$, and terms involving $X$ with non-positive subscript are not included in the model. 

Model (\ref{eqn:1}) is composed of two parts.
First, for the structural component, the mean model is specified by lag coefficients $\boldsymbol{\beta} = (\beta_0, ..., \beta_L)'$ and control mean $\mu$. The immediate treatment effect is measured by $\beta_0$, and the carryover effect due to treatment on $l$ days ago is measured by $\beta_l$ for $l>0$. In the model, we assume the carryover effect beyond day $L$ is zero. As such, the total carryover treatment effect is captured as
$$\delta \triangleq \sum_{l=1}^L\beta_l = E(Y_t|X_{t-1} = 1,...X_{t-L} = 1, X_t) - E(Y_t|X_{t-1} = 0,...,X_{t-L} = 0, X_t).$$
Hence, the model naturally breaks down total treatment effect into $\beta_0$ and $\delta$.

Second, for the stochastic component, temporal dependency between errors is specified  using an order-$p$ autoregressive error model with autoregression coefficient $\boldsymbol{\phi} = (\phi_1, ..., \phi_p)'$.
Let $B$ denote the backshift operator, that is,  having $\Phi(B)=1-\phi_{1}B-\phi_{2}B^{2}-...-\phi_pB^{p}$ so that the autoregression model for the error terms can be written as $\Phi(B)\epsilon_{t}=w_{t}$. It is often convenient to work with the transformed data $Y^*_t=\Phi(B)Y_t$ and $X_{t}^* = \Phi(B)X_{t}$ in the estimation steps.
Thus, applying $\Phi(B)$ to both sides of model (\ref{eqn:1}), we will rewrite the model 

\begin{equation} \label{transformed1}
    Y^*_{t}=\mu^*+\sum^{L}_{l=0}\beta_{l}X^*_{t-l}+w_t,
\end{equation}

for $t=p+1,...,n$, where  $\mu^* = \Phi(B)\mu$.
To stack the data in vector form, we have 

\begin{equation} \label{likelihood}
    (\boldsymbol{Y}^* \mid \boldsymbol{X}^*, \mu^*, \boldsymbol{\beta}) \sim N(\mu^*\boldsymbol{1}_{n-p}+\boldsymbol{X}^*\boldsymbol{\beta}, \sigma^2 \mathbf{I}_{n-p})
\end{equation}

where $\boldsymbol{Y}^* = (Y^*_{p+1},...,Y^*_{n})'$, $\boldsymbol{X}^*$ is a $(n-p)\times (L+1)$ matrix with $X^*_{k-l+p+1}$ being the $(k,l)$-th element of $\boldsymbol{X}^*$, $\boldsymbol{1}_{n-p}$ is a  1-vector of length $n-p$, and $\mathbf{I}_{n-p}$ is the identity matrix of dimension $n-p$. 
We denote $\tilde{\boldsymbol{\beta}} = (\mu, \boldsymbol{\beta}')'$ and $\tilde{\boldsymbol{X}^*} = (\Phi(B)\boldsymbol{1}_{n-p}, \boldsymbol{X}^*)$, so that $\tilde{\boldsymbol{X}^*}\tilde{\boldsymbol{\beta}} = \mu^*\boldsymbol{1}_{n-p}+\boldsymbol{X}^*\boldsymbol{\beta}$.

\subsection{Prior Distribution on the Mean Model}
\label{ss:prior}

We consider normal prior distribution for $\tilde{\boldsymbol{\beta}}$, that is, having

\begin{equation} \label{prior_beta_tilde}
    \tilde{\boldsymbol{\beta}} \sim N(\boldsymbol{0}, \sigma^2 \tilde{\boldsymbol{\Omega}}^{-1}),
\end{equation}

where $\tilde{\mathbf{\Omega}} = diag(c_0, \boldsymbol{\Omega})$ so that the prior variance of $\mu$ is $\sigma^2 c_0^{-1}$ and the prior variance-covariance matrix of $\boldsymbol{\beta}$ is $\sigma^2 \boldsymbol{\Omega}^{-1}$.
We note that the prior variance depends on the variance $\sigma^2$ of the observations: such dependence renders a fused ridge penalized estimation procedure that is free of the variance parameters, resulting in computational stability; see Equation (\ref{eq:penalty}) below.
We will postulate a non-informative prior on $\mu$ by setting $c_0$ to be a small number, and we will consider $\boldsymbol{\Omega}$ of the following form:

\begin{equation}
    \left(\begin{array}{cccccc}{\lambda_{0}+\lambda^*_{0}} & {-\lambda^*_{0}} & {0}  & {\ldots} & {\ldots} & {0} \\
    {-\lambda^*_{0}} & {\lambda_{1}+\lambda^*_{0}+\lambda^*_{1}} & {-\lambda^*_{1}} & {\ldots} & {\ldots} & {0} \\
    {0} & {-\lambda^*_{1}} & {\lambda_{2}+\lambda^*_{1}+\lambda^*_{2}} & {\ldots} & {\ldots} & {0} \\
    {\vdots} & {\vdots} & {\vdots} & {\ddots} & {\vdots} & {\vdots}\\
    {0} & {0} & {0}  & {\ldots} & {{\lambda_{L-1}}+\lambda^*_{L-2}+\lambda^*_{L-1}} & {-\lambda^*_{L-1}}\\
    {0} & {0} & {0}  & {\ldots} & {-\lambda^*_{L-1}} & {{\lambda_L}+\lambda^*_{L-1}+\lambda^*_{L}}\end{array}\right),
    \label{eq:omega}
\end{equation}

where the hyperparameters $\lambda_l, \lambda_l^* >0$, for $l=0, \ldots, L,$ are constrained to increase over $l$.
As a result of the monotonicity constraint, a lag coefficient $\beta_l$ at a  greater lag $l$ is associated with a larger diagonal element in precision $\boldsymbol{\Omega}$, thus shrinking $\beta_l$ toward the prior mean (zero) to a greater extent. This effectively  addresses collinearity of the lag coefficients without imposing strong parametric structure to $\boldsymbol{\beta}$.
In addition, using the normal prior \eqref{prior_beta_tilde} with precision matrix (\ref{eq:omega}), we can show that the maximum {\emph{ a posteriori}} probability estimate of  $\boldsymbol{\tilde{\beta}}$ minimizes a fused ridge-type penalty:

\begin{equation}
    (\boldsymbol{Y}^* - \tilde{\boldsymbol{X}}^*\tilde{\boldsymbol{\beta}})' (\boldsymbol{Y}^* - \tilde{\boldsymbol{X}}^* \tilde{\boldsymbol{\beta}}) + c_0\mu^2 + \sum_{l=0}^L \lambda_l\beta_l^2 + \sum_{l=0}^{L}\lambda^*_l(\beta_l-\beta_{l+1})^2
\label{eq:penalty}
\end{equation}

where $\beta_{L+1} \triangleq 0$, thus giving insights on how the proposed prior constrains the lag coefficients:
it regularizes not only the $\ell_2$-norm of the coefficients but also their successive differences, thereby enhancing local smoothness. The equivalence between the Bayesian inference and the fused ridge regularization (\ref{eq:penalty}) is proved
in Appendix A. 

There are many ways to specify $\lambda_l$ and $\lambda_l^*$ to meet the monotonicity constraints. In this article,  we consider $\lambda_l = \exp\{\gamma_1 (l+1) \}-1$ and $\lambda^*_l = \exp\{ \gamma_2 (l+1) \}-1$ for $\gamma_1, \gamma_2 >0$, so that $\gamma_1$ controls the rate at which the ridge penalty in (\ref{eq:penalty}) increases, and $\gamma_2$ controls the increasing rate of smoothness of the coefficient curve $\boldsymbol{\beta}$.
Instead of treating these hyperparameters as fixed, we postulate a truncated standard exponential hyperprior on  $(\gamma_1,\gamma_2)$, that is, having probability density function

\begin{equation}
    \pi(\gamma_1, \gamma_2) \propto \exp{(-\gamma_1-\gamma_2)}\boldsymbol{1}_{S_{\boldsymbol{\gamma}}}(\gamma_1,\gamma_2)
\end{equation}

where the support $S_{\boldsymbol{\gamma}}$ includes all pairs $(\gamma_1,\gamma_2)$ with which the precision matrix $\mathbf{\Omega}$ is positive definite.  As such, the degree of ridge and smooth penalization can be 
determined according to the  posterior distribution of the pair.

\subsection{Prior Distribution on the Error Model}
\label{ss:prior2}

We put the Jeffreys prior for the error variance $\sigma^2$, that is, having density function

\begin{equation}
\label{prior_sigma2}
    \pi(\sigma^2) \propto 1/\sigma^2
\end{equation}

Note that any inverse-gamma prior for $\sigma^2$ would maintain conjugacy, and the Jeffreys prior can be regarded as an improper limit of inverse-gamma prior distribution. 

For the autoregressive process, we consider a truncated normal prior for $\boldsymbol{\phi}$ subject to the constraint that the error process is stationary.
Specifically, we postulate

\begin{equation}
    \boldsymbol{\phi} \sim N_{p}\left(0_{p},200 \times \mathbf{I}_{p} \right)\boldsymbol{1}_{S_{\boldsymbol{\phi}}}(\boldsymbol{\phi})
\label{eq:phi}
\end{equation}

where ${S_{\boldsymbol{\phi}}}(\boldsymbol{\phi})$ denotes the support where all roots of the polynomial $\Phi(B) = 1-\sum_{l=1}^p\phi_l B^l$ are outside the unit circle. The process  $\{\epsilon_t: t=1, 2, \ldots\}$ is stationary when $\boldsymbol{\phi} \in S_{\boldsymbol{\phi}}(\boldsymbol{\phi})$. \cite{chib1993bayes} Note that the range of each $\phi_l$ is $(-1,1)$; thus, a prior variance of 200 in (\ref{eq:phi}) essentially amounts to a flat prior.

\section{Conditional Posterior Distributions}
\label{s:posterior}

The proposed Bayesian model includes several conditionally conjugate priors, which facilitate posterior computations via a hybrid Metropolis-Hastings/Gibbs algorithm. We describe the conditional posterior distributions in this section.

Working with the likelihood (\ref{likelihood}) based on the transformed data $Y_t^*$, we obtain that $\tilde{\boldsymbol{\beta}}$ is conditionally normally distributed {\it a posteriori}:

\begin{equation} \label{posterior_beta}
  \tilde{\boldsymbol{\beta}} \mid \boldsymbol{Y},\tilde{\boldsymbol{X}},\sigma^2,\boldsymbol{\phi},\boldsymbol{\gamma} \sim N_{L+1}\left\{[\tilde{\boldsymbol{X}}^{*'}\tilde{\boldsymbol{X}}^* + \tilde{\boldsymbol{\Omega}}(\boldsymbol{\gamma})]^{-1}\tilde{\boldsymbol{X}}^{*'}\boldsymbol{Y}^*, \sigma^2[\tilde{\boldsymbol{X}}^{*'}\tilde{\boldsymbol{X}}^* + \tilde{\boldsymbol{\Omega}}(\boldsymbol{\gamma})]^{-1}  \right\}
\end{equation}

and that $\sigma^2$ has an inverse-gamma conditional posterior:

\begin{equation} \label{posterior_sigma2}
\sigma^2 \mid \boldsymbol{Y},\tilde{\boldsymbol{X}},\tilde{\boldsymbol{\beta}},\boldsymbol{\phi},\boldsymbol{\gamma} \sim \text{IG}\left[\frac{n-p+L+1}{2}, \frac{(\boldsymbol{Y}^*-\tilde{\boldsymbol{X}}^*\tilde{\boldsymbol{\beta}})'(\boldsymbol{Y}^* - \tilde{\boldsymbol{X}}^*\tilde{\boldsymbol{\beta}}) + \tilde{\boldsymbol{\beta}}'\tilde{\boldsymbol{\Omega}}(\boldsymbol{\gamma})\tilde{\boldsymbol{\beta}}}{2}\right]
\end{equation}

Note that the dependence of (\ref{posterior_beta}) and (\ref{posterior_sigma2}) on $\boldsymbol{\phi}$ is via the transformed data $\boldsymbol{Y^*}$.

Working with model \eqref{eqn:2} and \eqref{eq:phi}, we obtain the conditional posterior distribution of $\boldsymbol{\phi}$ is truncated multivariate normal: 

\begin{equation} \label{posterior_phi}
  \boldsymbol{\phi} \mid \boldsymbol{Y},\tilde{\boldsymbol{X}},\tilde{\boldsymbol{\beta}},\sigma^2,\boldsymbol{\gamma} \sim N_{p}\left[\left(\sigma^{-2}\boldsymbol{E}^{*'}\boldsymbol{E}^*+\sigma_{\boldsymbol{\phi}}^{-2}\mathbf{I}\right)^{-1}\sigma^{-2}\boldsymbol{E}^{*'}\boldsymbol{\epsilon}^*, \left(\sigma^{-2}\boldsymbol{E}^{*'}\boldsymbol{E}^{*}+\sigma_{\boldsymbol{\phi}}^{-2}\mathbf{I}\right)^{-1}  \right]\boldsymbol{1}_{S_{\boldsymbol{\phi}}}(\boldsymbol{\phi})
\end{equation}

where $\boldsymbol{\epsilon}^* = (\epsilon_{p+1}^*,\ldots, \epsilon_n^*)'$, $\epsilon_t^* = Y_t - \mu - \sum^{L}_{l=0}\beta_{l}X_{t-l}$, and $\boldsymbol{E}^*$ is a $(n-p)\times p$ matrix with $\epsilon_{p+k-j}^*$ being the $(k, j)$-th element. Because of conjugacy, the parameters $\tilde{\boldsymbol{\beta}}, \sigma^2,$ and $\boldsymbol{\phi}$ can be easily updated
in a Gibbs sampling fashion.

Using the likelihood (\ref{likelihood}) and prior of $\boldsymbol{\gamma}$ and $\tilde{\boldsymbol{\beta}}$, the conditional posterior distribution can be expressed as 

\begin{equation} \label{posterior_gamma}
    \pi(\boldsymbol{\gamma} \mid \boldsymbol{Y},\tilde{\boldsymbol{X}},\tilde{\boldsymbol{\beta}},\boldsymbol{\phi}, \sigma^2) \propto  |\sigma^{-2} \tilde{\boldsymbol{\Omega}}(\boldsymbol{\gamma})|^{\frac{1}{2}}\exp \left[-\frac{1}{2\sigma^2}\tilde{\boldsymbol{\beta}}'\tilde{\boldsymbol{\Omega}}(\boldsymbol{\gamma})\tilde{\boldsymbol{\beta}}\right]\exp(-\gamma_1-\gamma_2)\boldsymbol{1}_{S_{\boldsymbol{\gamma}}}(\gamma_1,\gamma_2).
\end{equation}

We propose to sample $\boldsymbol{\gamma}$ using a Metropolis-Hastings (MH) step with a uniform  $U(-a,a)$ proposal distribution,
that is, having an updating step $\gamma_{i, new} = \gamma_i + U(-a,a)$, where the tuning parameter $a$ is chosen such that the acceptance rate of proposed sample is around 50\%. \cite{gelman1996efficient} Note that updating the hyperparameter $\boldsymbol\gamma$ involves the calculation of the precision matrix $\tilde{\boldsymbol{\Omega}}(\boldsymbol{\gamma})$, which needs to be positive definite. The $(L+2) \times (L+2)$ precision matrix $\tilde{\boldsymbol{\Omega}}(\boldsymbol{\gamma})$ is a special case of tridiagonal matrix. The computational cost of regular algorithms for checking whether a given matrix is positive definite is at most $O(L^3)$. In this article, we implement an efficient computation algorithm with cost of $O(L)$.
Specifically, define an $(L+2)$-dimensional vector $\mathbf{c}=(c_0,c_2,...,c_{L+1})$ by

\begin{eqnarray*}
c_l = \begin{cases}
\lambda_0+\lambda^*_0, &~~l = 0,\\
(\lambda_l+\lambda^*_l)-\frac{1}{c_{l-1}}, &~~l= 1,2,...,L+1
\end{cases}
\label{eq0}
\end{eqnarray*}

El-Mikkawy showed that the $\tilde{\boldsymbol{\Omega}}(\boldsymbol{\gamma})$ is positive definite if and only if $c_l>0$ for $l=0,1,...L+1$. Thus, the problem boils down to checking the positiveness of elements in $\mathbf{c}$.\cite{el2004fast} The complete algorithm is summarized in Appendix B.

\section{Simulation study}
\label{s:simulation}

\subsection{Comparison Methods}
In this section, we evaluate the performance of the proposed BDLM-AR using simulation studies.  At the end of each simulated trial, we fitted BDLM-AR with lag $L=7$ and AR(1), that is, having

\begin{equation}
    Y_t= \mu + \sum^7_{l=0}\beta_lX_{t-l}+\epsilon_t
\label{eq:simmodel}
\end{equation}

where $\epsilon_t = \phi \epsilon_{t-1}+w_t$ and $w_t \sim N(0, \sigma^2)$.
Posterior distributions were derived using the hybrid Metropolis Hastings/Gibbs algorithm described in the previous section with 50,000 iterations, a burn-in period of 25,000, and $a = 0.2$ for sampling $\gamma$ in the MH step.

We compared BDLM-AR with some existing methods including the Bayesian distributed lag model (BDLagM),\cite{welty2009bayesian} Bayesian ridge DLM (BR-DLM) with a mean zero normal prior for $\tilde{\boldsymbol{\beta}}$, and a non-informative prior Bayesian DLM (NB-DLM) with a flat improper priors on each parameter in $\tilde{\boldsymbol{\beta}}$. 
These existing methods would use the same mean model (\ref{eq:simmodel}) but assume independent errors without accounting for autocorrelation.

In addition, as a benchmark, we include the parametric Koyck's DLM \cite{koyck1954distributed} which assumes the knowledge of the true autoregressive coefficients. For Bayesian models, we estimate the parameters using the posterior means and for Koyck model, we use the maximum likelihood estimates.

\subsection{Simulation Scenarios and Data Generation}

In each simulated N-of-1 trial,  measurements were collected daily for 120 days, under one of two possible treatment sequences.  In the first sequence, a participant would receive
$x_t =1$ on the first 30 days and the last 30 days, and receive $x_t=0$ between days 31 and 90; that is,

\begin{equation*}
x_t^{(1)} =
\begin{cases}
1 & t = 30s+1,...,30s+30 \text{ for } s = 0, 3\\
0 & t = 30s+1,...,30s+30 \text{ for } s = 1, 2.
\end{cases}
\end{equation*}
In the second treatment sequence, a participant would switch treatments more frequently; specifically, 
\begin{equation*}
x_t^{(2)}=
\begin{cases}
1 & t = 15s+1,...,15s+15 \text{ for } s = 0, 3, 5, 6\\
0 & t = 15s+1,...,15s+15 \text{ for } s = 1, 2, 4, 7.
\end{cases}
\end{equation*}

For each  treatment sequence, the data were generated according to model (\ref{eq:simmodel}) under five sets of lag coefficients (lag curves, LC):

\begin{enumerate}[wide = 12pt]
{
    \item [LC1.] Exponential decay curve: $\boldsymbol{\beta} = (5, 2.5, 1.25, 0.625, 0.3125, 0, 0, 0)^{'}$;
    \item [LC2.] Exponential decay curve with oscillation: $\boldsymbol{\beta} = (5, 2.5, -1.25, -0.625, 0.3125, 0, 0, 0)^{'}$;
    \item [LC3.] Slow absorption curve: $\boldsymbol{\beta} = (1.51, 2.75, 3.36, 2.03, 0.34, 0, 0, 0)^{'}$;
    \item [LC4.] Slow absorption curve with oscillation: $\boldsymbol\beta =(1.51, 2.75, -3.36, -2.03, 0.34, 0, 0, 0)^{'}$;
    \item [LC5.] No carryover effect: $\boldsymbol{\beta}=(10, 0, 0, 0, 0, 0, 0, 0)^{'}$.
    }
\end{enumerate}

The exponential decay curves (LC1 and LC2) specify coefficients that diminish in magnitude as lag lengthens. Specifically, the coefficients under LC1 decrease geometrically, which is aligned with the assumption of Koyck DLM. The slow absorption curves (LC3 and LC4) reflect scenarios where the carryover effect peaks at day 2 after treatment. LC5 is the null scenario where there is no carryover effect. 
The magnitudes of the coefficients were chosen in these scenarios so that $\|\boldsymbol\beta\|_1\approx 10$; in addition, the total carryover effects ($\delta$) are 4.69, 0.94, 8.48, -2.30 and 0 respectively for LC1-LC5. We consider $\sigma = 10, 20$ and $\phi = 0.5, 0.2$ for the stochastic component in data generation. For each of these scenarios, the methods were evaluated using 100 simulated trials.

\subsection{Simulation Results}

\begin{figure}[hp]
\centering
\includegraphics[width=1.0\linewidth]{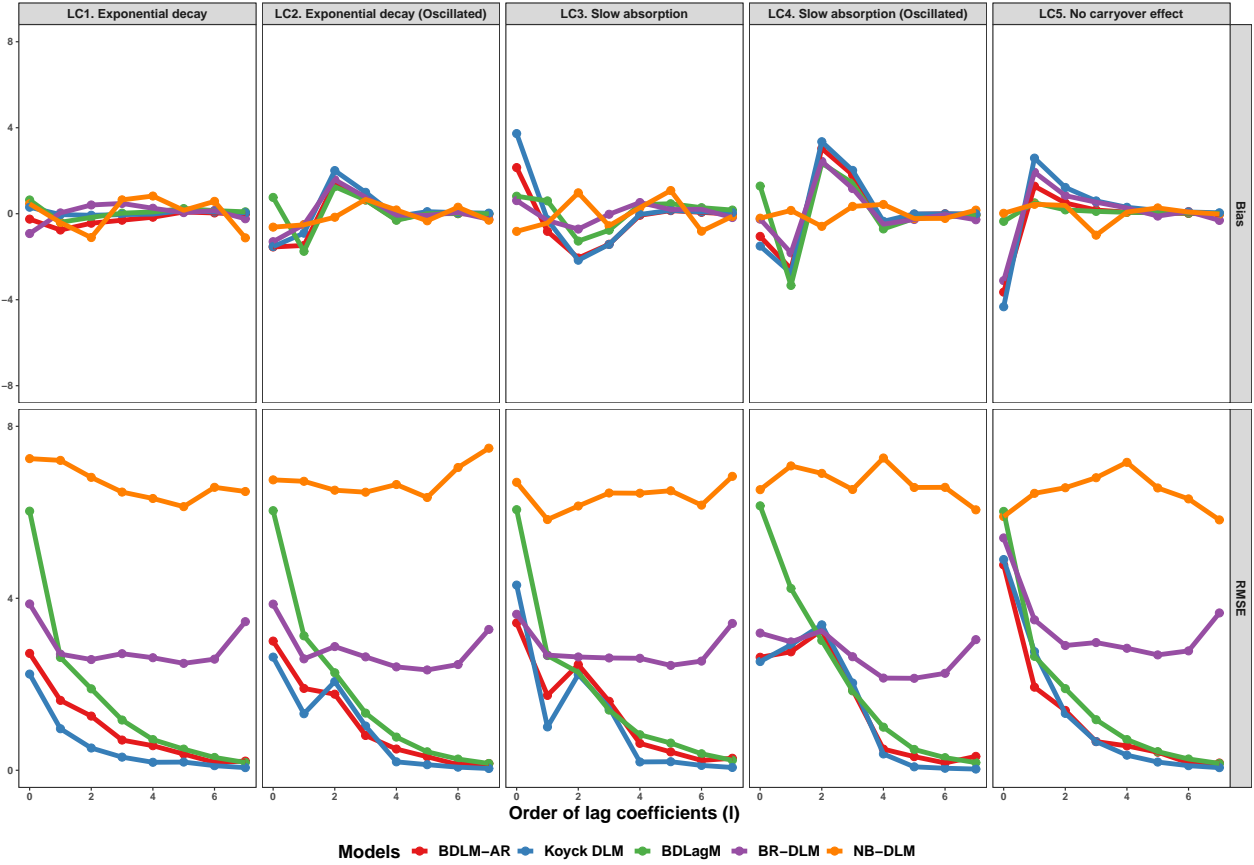}
\caption{Bias and RMSE of lag coefficients estimates under five lag curves, treatment sequence $x_t^{(1)}$, $\sigma$ = 10, $\phi$ = 0.5.
\label{fig:confidenceBandLinePlot}}
\end{figure}

Figure \ref{fig:confidenceBandLinePlot} shows the bias and root mean squared error (RMSE) of the posterior means of individual lag coefficients. As expected, the biases of NB-DLM are relatively small; however, the method also has the largest RMSE uniformly because of the use of non-informative prior.  The biases of the other methods are comparable, and are generally small compared to RMSE. While the $\ell_2$ penalty in BR-DLM on the lag coefficients reduces variability when compared to NB-DLM, the additional constraints on diminishing coefficients imposed by BDLagM and the proposed BDLM-AR further reduce RMSE for large lag $l$.
Additionally, since the proposed BDLM-AR  explicitly incorporates ridge-type regularization on the lag coefficients, it results in smaller RMSE for $\beta_0$ and the earlier lag coefficients (e.g. $\beta_1$).
However, as a trade-off, the bias of BDLM-AR for early lag coefficients will be slightly inflated as compared to BDLagM, BR-DLM and NB-DLM, especially when true lag coefficient curve has frequent fluctuation. 
The benchmark method, Koyck DLM, performs best in the exponential decay case, where the true coefficient of autoregressive error ($\phi$) is assumed to be known. The proposed BDLM-AR has very similar performance as Koyck DLM. Note that the coefficient of autoregressive error is estimated directly from the proposed BDLM-AR model, which is more practical in real application.
In summary, the proposed BDLM-AR generally results in smallest RMSE for all lag coefficients.

\begin{figure}[hp]
\begin{center}
\includegraphics[width=1.0\linewidth]{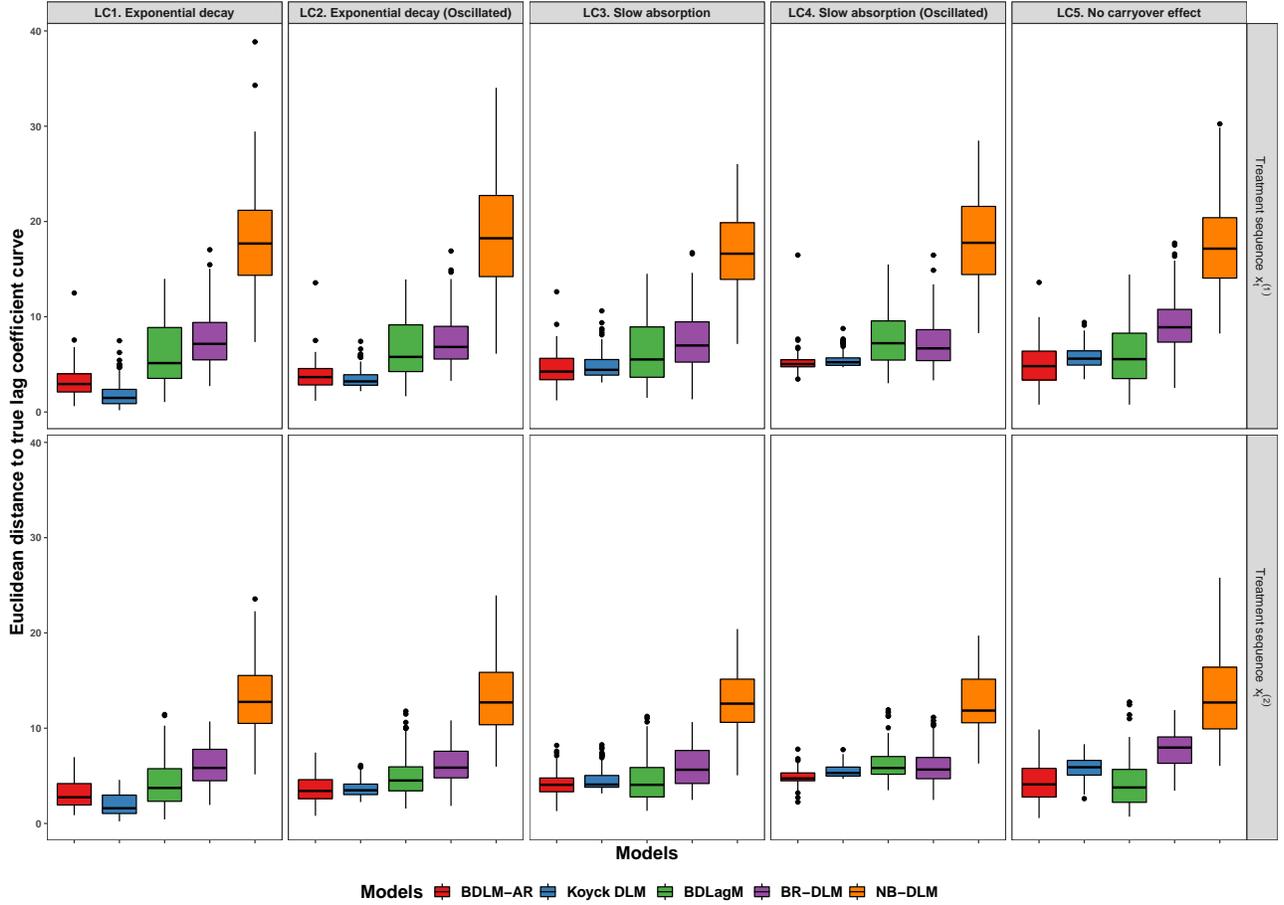}   
\end{center}
\caption{Euclidean distance to true lag curves under: Treatment sequence $x_t^{(1)}$ vs. Treatment sequence $x_t^{(2)}$.
\label{fig:distanceUnderDesign}}
\end{figure}

To further examine the performance of each method in estimating the lag curve in aggregate, Figure \ref{fig:distanceUnderDesign} (top panel) plots the Euclidean distance between the vector of estimated lag coefficients and the vector of true lag coefficients.
Under LC1 (exponential decay), the Koyck DLM has the best performance overall. This is not surprising because (i) the Koyck model mimics the coefficients under LC1 closely, and (ii) Koyck DLM assumes knowledge of the true autoregressive coefficients used in the simulation and hence it is not a method that can be implemented in practice. Thus, this comparison serves as a benchmark about the efficiency of the proposed BDLM-AR and other methods. The figure demonstrates that the proposed BDLM-AR produces smaller distance from the true lag curve than the other methods. 

Table 1 gives the bias and RMSE in the estimation of the total effect ($\sum_{l=0}^7\beta_l$), the total carryover effect ($\delta = \sum_{l=i}^7\beta_l$) and the immediate effect ($\beta_0$) under different lag curves with $\sigma = 10$ and $\phi = 0.5$ under treatment sequence $x_t^{(1)}$.
Results for other values of $\phi$, $\sigma$ and treatment sequence are similar and are provided in Figure A1 to A3 in the online Supporting Information.
Overall, the proposed BDLM-AR yields consistently lower RMSE in estimating total effect, carryover effects, and immediate effects than other comparison methods, except for Koyck DLM. This is consistent with what we observe in Figures \ref{fig:confidenceBandLinePlot} and \ref{fig:distanceUnderDesign}.
We note that the advantages of BDLM-AR in terms of RMSE for the total carryover effect $(\delta)$ and immediate effect $(\beta_0)$ are more pronounced than that for total effect ($\delta + \beta_0$).  This is indeed the motivation that we set out to accomplish: to decompose the treatment effects and separate carryover effect from the immediate effect.

\begin{table}[hp]
\caption{\quad Summary of evaluation metrics (best values in bold) of total effect, total carryover effect and immediate effect($\beta_0$) under five lag curves, treatment sequence $x_t^{(1)}$, $\sigma$ = 10 and $\phi$ = 0.5. 
\label{table:metricsScenarios}}
\centering
\resizebox{\linewidth}{!}{%
\begin{tabular}{clcccccc}
\hline
                       &                                       & Truth & BDLM-AR        & Koyck DLM     & BDLagM         & BR-DLM        & NB-DLM         \\ \hline
\multirow{18}{*}{Bias} & \textbf{Total Effect}                          &       &                &               &                &               &                \\
                       & \qquad LC1. Exponential decay                & 10    & -1.82          & 0.41          & 0.65           & 0.21          & \textbf{0.02}  \\
                       & \qquad LC2. Exponential decay (Oscillated)   & 5.94    & -1.24          & 0.61          & 0.61           & \textbf{0.20} & -0.79          \\
                       & \qquad LC3. Slow absorption                  & 10    & -2.04          & \textbf{0.12} & 0.71           & 0.30          & -0.42          \\
                       & \qquad LC4.   Slow absorption (Oscillated)   & -0.79    & 0.73           & 0.73          & 0.59           & 0.46          & \textbf{-0.15} \\
                       & \qquad LC5. No carryover                     & 10    & -1.60          & 0.65          & 0.58           & \textbf{0.14} & 0.25           \\
                       & \textbf{Total Carryover Effect}                &       &                &               &                &               &                \\
                       & \qquad LC1. Exponential decay                & 4.69  & -1.57          & 0.10          & \textbf{0.01}  & 1.13          & -0.44          \\
                       & \qquad LC2.   Exponential decay (Oscillated) & 0.94  & 0.31           & 2.14          & \textbf{-0.15} & 1.50          & -0.17          \\
                       & \qquad LC3. Slow absorption                  & 8.48  & -4.19          & -3.61         & \textbf{-0.11} & -0.31         & 0.40           \\
                       & \qquad LC4.   Slow absorption (Oscillated)   & -2.30 & 1.78           & 2.23          & -0.70          & 0.71          & \textbf{0.06}  \\
                       & \qquad LC5. No carryover                     & 0     & 2.05           & 4.98          & 0.94           & 3.25          & \textbf{0.23}  \\
                       & \textbf{Immediate Effect}                      &       &                &               &                &               &                \\
                       & \qquad LC1. Exponential decay                & 5     & \textbf{-0.25} & 0.30          & 0.64           & -0.92         & 0.46           \\
                       & \qquad LC2.   Exponential decay (Oscillated) & 5     & -1.55          & -1.53         & 0.76           & -1.30         & \textbf{-0.62} \\
                       & \qquad LC3. Slow absorption                  & 1.51  & 2.15           & 3.73          & 0.81           & \textbf{0.61} & -0.82          \\
                       & \qquad LC4.   Slow absorption (Oscillated)   & 1.51  & -1.05          & -1.51         & 1.29           & -0.25         & \textbf{-0.21} \\
                       & \qquad LC5. No carryover                     & 10    & -3.65          & -4.33         & -0.35          & -3.11         & \textbf{0.02}  \\ \hline
\multirow{18}{*}{RMSE} & \textbf{Total Effect}                          &       &                &               &                &               &                \\
                       & \qquad LC1. Exponential decay                & 10    & \textbf{3.61}  & 3.87          & 4.05           & 4.17          & 3.96           \\
                       & \qquad LC2.   Exponential decay (Oscillated) & 5.94  & \textbf{2.95}  & 3.90          & 4.05           & 4.11          & 3.93           \\
                       & \qquad LC3. Slow absorption                  & 10    & \textbf{3.81}  & 3.85          & 4.06           & 4.16          & 4.40           \\
                       & \qquad LC4.   Slow absorption (Oscillated)   & -0.79 & \textbf{2.28}  & 3.93          & 4.04           & 4.01          & 3.99           \\
                       & \qquad LC5. No carryover                     & 10    & \textbf{3.54}  & 3.91          & 4.04           & 4.18          & 3.93           \\
                       & \textbf{Total Carryover Effect}                &       &                &               &                &               &                \\
                       & \qquad LC1. Exponential decay                & 4.69  & 3.03           & \textbf{2.01} & 6.57           & 4.83          & 7.61           \\
                       & \qquad LC2.   Exponential decay (Oscillated) & 0.94  & \textbf{2.25}  & 2.87          & 6.56           & 4.83          & 6.31           \\
                       & \qquad LC3. Slow absorption                  & 8.48  & 5.13           & \textbf{4.15} & 6.58           & 4.64          & 6.94           \\
                       & \qquad LC4.   Slow absorption (Oscillated)   & -2.30 & 2.96           & \textbf{2.94} & 6.62           & 4.44          & 6.39           \\
                       & \qquad LC5. No carryover                     & 0     & \textbf{3.39}  & 5.36          & 6.65           & 6.06          & 6.56           \\
                       & \textbf{Immediate Effect}                      &       &                &               &                &               &                \\
                       & \qquad LC1. Exponential decay                & 5     & 2.71           & \textbf{2.23} & 6.00           & 3.85          & 7.21           \\
                       & \qquad LC2.   Exponential decay (Oscillated) & 5     & 2.99           & \textbf{2.62} & 6.01           & 3.85          & 6.72           \\
                       & \qquad LC3. Slow absorption                  & 1.51  & \textbf{3.42}  & 4.30          & 6.03           & 3.61          & 6.67           \\
                       & \qquad LC4.   Slow absorption (Oscillated)   & 1.51  & 2.62           & \textbf{2.52} & 6.12           & 3.18          & 6.49           \\
                       & \qquad LC5. No carryover                     & 10    & \textbf{4.77}  & 4.90          & 5.99           & 5.39          & 5.87           \\ \hline
\end{tabular}}
% \vspace*{1cm}
\end{table}

\subsection{Effects of Design}

Figure \ref{fig:distanceUnderDesign} shows that the Euclidean distance between the vector of estimated lag coefficients and the truth under $x_t^{(2)}$ (bottom panel) is smaller than that under $x_{t}^{(1)}$ (top panel) suggesting frequently switching treatments will help improve the information content in N-of-1 trial data.  This is in line with what we expect because collinearity of exposure lags will be lessened as treatments change frequently, while the total duration is held fixed.  An implication in practice is that we should alternate intervention as frequent as it is feasible.
The relative performance among methods is the same regardless of the treatment sequence, that is, the proposed BDLM-AR yields the shortest distance from the true lag coefficients  $\boldsymbol{\beta}$.

\subsection{Effects of Model Misspecification}

In the previous subsections, BDLM-AR and other methods use a working mean model with $L=7$ and an AR(1) model for autoregressive errors. These working models correctly specify (or include the data generation model) in the previous simulation study.  
In this subsection, we investigate the robustness of BDLM-AR under model mis-specifications.  Specifically, we will consider 
(1) the working mean model with $L = 0, 1, \ldots, 7$;
(2) the stochastic components that assume autoregressive error order of $p = 0, 1, 7$.  
That is, we consider a total of 24 BDLM-AR models. 

In data generation, we use LC1 as the true mean model, where $\beta_l >0$ for $l = 0,1, 2,3, 4$, and we consider true scenarios for the errors:

\begin{enumerate}[wide = 12pt]
\item[E1.] AR(1) with $\phi = 0.5$;
\item[E2.] Autoregressive model with $\phi_1=0.5, \phi_2 = 0,\phi_3 = 0, \phi_4 = 0.3, \phi_5 = 0, \phi_6 = 0.2$.
\end{enumerate}

Note that, under the scenario LC1/E1, a working model with $L<4$ or $p=0$ under-specifies the true model. Likewise, under L1/E2, a working model with $L<4$ or $p=0,1$ under-specifies the true model.

Table \ref{table:misspecification} summarizes the RMSE of these 24 models under the two scenarios (LC1/E1 and LC/E2) with $\sigma = 10$ under $x_{t}^{(1)}$.
It can be seen that misspecified lag length has little influence on estimating total effect, total carryover effect and immediate effect, while under-specified error AR order will increase RMSE of parameters to a higher level than over-specified error AR order.
Note that when choosing a small lag length value, we can hardly acquire estimation about the whole DL curve, as well as the information on the duration of carryover effect. Therefore, when the lag length is unknown, we suggest to fit data with a reasonable long lag length. For error autoregressive order, when the true orders are unknown, it is also suggested to fit a model with high autoregressive order.

\begin{table}[hp]
\caption{\quad Summary of RMSE of total effect, total carryover effect and immediate effect($\beta_0$) fitted using BDLM-AR model with different lag length and error autoregressive order. LC1.Exponential decay curve, treatment sequence $x_t^{(1)}$, $\sigma$ = 10, E1. AR(1) with $\phi = 0.5$ and E2. Autoregressive model with $\phi_1=0.5, \phi_2 = 0,\phi_3 = 0, \phi_4 = 0.3, \phi_5 = 0, \phi_6 = 0.2$ are used to generate simulated data.
\label{table:misspecification}}
\centering
\resizebox{\linewidth}{!}{%
\begin{tabular}{ccccccccc}
\hline
                                        &     & \multicolumn{3}{c}{Truth for errors: E1} &  & \multicolumn{3}{c}{Truth for errors: E2} \\ \hline
                                        & Lag & AR(7)        & AR(1)       & AR(0)       &  & AR(7)        & AR(1)       & AR(0)       \\ \cline{3-5}  \cline{7-9} 
\multirow{8}{*}{Total Effect}           & 7   & 4.32         & 3.85        & 3.95        &  & 5.85         & 8.32        & 10.72       \\
                                        & 6   & 4.37         & 3.88        & 3.92        &  & 5.92         & 8.25        & 10.62       \\
                                        & 5   & 4.44         & 3.91        & 3.93        &  & 5.89         & 8.21        & 10.55       \\
                                        & 4   & 4.49         & 3.98        & 3.94        &  & 6.01         & 8.06        & 10.48       \\
                                        & 3   & 4.59         & 4.02        & 3.93        &  & 6.02         & 7.81        & 10.40       \\
                                        & 2   & 4.63         & 4.07        & 3.91        &  & 6.08         & 7.68        & 10.34       \\
                                        & 1   & 4.75         & 4.15        & 3.87        &  & 6.12         & 7.43        & 10.26       \\
                                        & 0   & 4.01         & 3.71        & 3.77        &  & 5.82         & 8.58        & 10.52       \\ \hline
\multirow{8}{*}{Total Carryover Effect} & 7   & 3.56         & 3.18        & 3.85        &  & 3.55         & 5.24        & 6.69        \\
                                        & 6   & 3.51         & 3.17        & 3.73        &  & 3.60         & 5.15        & 6.38        \\
                                        & 5   & 3.42         & 3.15        & 3.61        &  & 3.60         & 4.95        & 6.08        \\
                                        & 4   & 3.41         & 3.12        & 3.49        &  & 3.73         & 4.67        & 5.68        \\
                                        & 3   & 3.41         & 3.12        & 3.36        &  & 3.70         & 4.29        & 5.33        \\
                                        & 2   & 3.38         & 3.10        & 3.15        &  & 3.73         & 4.00        & 4.62        \\
                                        & 1   & 3.58         & 3.33        & 2.92        &  & 3.86         & 3.88        & 3.54        \\
                                        & 0   & -            & -           & -           &  & -            & -           & -           \\ \hline
\multirow{8}{*}{Immediate Effect}       & 7   & 3.09         & 2.91        & 3.54        &  & 3.61         & 4.54        & 8.52        \\
                                        & 6   & 3.02         & 2.91        & 3.47        &  & 3.56         & 4.50        & 8.46        \\
                                        & 5   & 3.03         & 2.87        & 3.38        &  & 3.58         & 4.52        & 8.40        \\
                                        & 4   & 3.04         & 2.85        & 3.31        &  & 3.57         & 4.54        & 8.29        \\
                                        & 3   & 3.03         & 2.88        & 3.28        &  & 3.54         & 4.66        & 8.25        \\
                                        & 2   & 3.09         & 2.97        & 3.29        &  & 3.64         & 4.91        & 8.28        \\
                                        & 1   & 3.15         & 3.15        & 3.49        &  & 3.71         & 5.21        & 8.48        \\
                                        & 0   & 5.39         & 5.62        & 6.00        &  & 6.35         & 9.02        & 11.75       \\ \hline
\end{tabular}}
% \vspace*{1cm}
\end{table}

\section{Application to Light Therapy Study}
\label{s:application}

The data set we used is from the light therapy study,\cite{kronish2020clinical} which studies the effectiveness of bright white light therapy for depressive symptoms within cancer survivors. Besides bright white intervention (10,000 lux), dim red (50 lux) was used as a control intervention, which lacks sufficient light intensity to affect cells from retina. Patients were instructed to use one of two portable lightboxes each morning for 30 minutes per day. For each patient, the whole study duration was 12 weeks. One intervention was assigned on the first three weeks and last three weeks and the other intervention was assigned between the fourth week and the ninth week. The initial intervention was randomized, either bright white lightbox or dim red lightbox. The collected outcomes were depressive symptom and fatigue symptom, which were tracked using a smartphone application. The outcomes were measured by patient's self-reported standard single-item visual analog scale from 0-not at all depressed/tired to 10-extremely depressed/tired.
Some occasional missing outcomes were imputed using average non-missing value in the corresponding treatment block.

We fit the data with the proposed BDLM-AR model with $L=7$. Two autoregressive order of BDLM-AR model AR(1) and AR(7) were used. Convergence of all the MCMC were checked using both trace plots and the Gelman–Rubin diagnostics. \cite{gelman1992inference} To be specific, the potential scale reduction factor for lag coefficients, autoregressive coefficients and model variance all are smaller than 1.2, indicating the convergence of posterior samples. \cite{brooks1998general} In addition to the comparison Bayesian distributed lag models, we also fit frequentist autoregressive regression models (RegAR) with $p=1$ and 7.

Table \ref{table:twoSubjectResults} shows the posterior means of the coefficients by the Bayesian methods and the maximum likelihood estimates by RegAR using depressive symptom outcome.
For patient 7706, the RegAR and other Bayesian DLM models indicate a weak insignificant total effect of bright white intervention in relieving depressive symptom.
However, BDLM-AR(7) model identifies a significant strong effect of bright white intervention as -0.52 (90\% CI: -1.07, -0.02). Due to the adjustment on outcome serial correlation and ridge-type penalty on DL coefficients, the credible intervals of DL coefficients are much smaller than those estimated from models using white noise. To check the fitness of each model, we used Ljung–Box test to examine autocorrelation of the residuals,\cite{ljung1978measure} and the corresponding p-values of $\chi^2$-test are also shown in Table \ref{table:twoSubjectResults}. No statistically significant autocorrelation is found in residuals of BDLM-AR model. We also found a second peak of treatment effect within patient 7706 two days after the immediate effect. For patient 7708, we observe a similar estimation between different models in terms of total effect. Treatment total effect estimated from BDLM-AR(7) model is -1.13 (90\% CI: -3.14, 0.28), which is slightly lower than that from other models. Extra information obtained from BDLM-AR model is that the majority of treatment effect lasts for around two days. For RegAR(1) model, BDLagM, BR-DLM and NB-DLM, statistically significant autocorrelation is found in residuals, indicating an inadequacy of model fitting. Analysis results using fatigue symptom outcome can be found in Table A1 in the online Supporting Information.

\begin{table}[hp]
\caption{\quad Lag coefficient estimates for light therapy study using depressive symptom outcome. 90\% credible intervals/confidence intervals are in brackets. P-value of Ljung-Box test for each model is on the last row.
\label{table:twoSubjectResults}}
\centering
\resizebox{\linewidth}{!}{%
\begin{tabular}{cccccccc}
\hline \hline
                                           & \multicolumn{7}{c}{Subject ID: 7706}                                                                                                                        \\ \cline{2-8} 
                                           & BDLM-AR(1)           & BDLM-AR(7)           & RegAR(1)             & RegAR(7)              & BDLagM               & BR-DLM              & NB-DLM              \\ \hline
\multicolumn{1}{c}{$\mu$}                  & 2.56 (2.20,2.91)   & 2.02 (0.07,2.88)    & 2.58 (2.21,2.95)   & 3.14 (2.05,4.23)   & 2.56 (2.34,2.79)   & 2.57 (2.36,2.77)   & 2.58 (2.35,2.80)   \\
\multicolumn{1}{c}{Total effect}           & -0.01 (-0.39,0.38) & -0.52 (-1.07,-0.02) & -0.02 (-0.52,0.49) & -0.36 (-0.93,0.21) & 0.04 (-0.31,0.39)  & 0.02 (-0.29,0.35)  & 0 (-0.38,0.38)     \\
\multicolumn{1}{c}{Total carryover effect} & 0.02 (-0.23,0.34)  & -0.22 (-0.74,0.14)  & -                  &                   & 0.28 (-0.52,1.07)  & 0.09 (-0.32,0.55)  & 0.24 (-0.62,1.10)  \\
\multicolumn{1}{c}{$\beta_0$}              & -0.03 (-0.41,0.32) & -0.30 (-0.72,0.08)  & -0.02 (-0.52,0.49) & -0.36 (-0.93,0.21) & -0.23 (-1.02,0.56) & -0.07 (-0.47,0.26)  & -0.24 (-1.10,0.61) \\
\multicolumn{1}{c}{$\beta_1$}              & 0.01 (-0.20,0.22)  & -0.01 (-0.30,0.37)  & -                  & -                 & 0.05 (-0.86,0.97)  & -0.02 (-0.43,0.37)  & -0.02 (-1.20,1.16) \\
\multicolumn{1}{c}{$\beta_2$}              & 0.01 (-0.11,0.15)  & -0.08 (-0.35,0.10)  & -                  & -                 & 0.10 (-0.30,0.49)  & 0.03 (-0.37,0.46)  & 0.09 (-1.10,1.28)  \\
\multicolumn{1}{c}{$\beta_3$}              & 0.01 (-0.07,0.10)  & -0.08 (-0.34,0.05)  & -                  & -                 & 0.05 (-0.14,0.24)  & 0.06 (-0.33,0.51)  & 0 (-1.16,1.19)     \\
\multicolumn{1}{c}{$\beta_4$}              & 0.01 (-0.03,0.07)  & -0.03 (-0.21,0.07)  & -                  & -                 & 0.03 (-0.08,0.14)  & 0.13 (-0.22,0.69)  & 0.90 (-0.27,2.07)  \\
\multicolumn{1}{c}{$\beta_5$}              & 0 (-0.04,0.03)     & -0.02 (-0.15,0.06)  & -                  & -                 & 0.02 (-0.05,0.09)  & -0.10 (-0.64,0.26)  & -0.90 (-2.08,0.27) \\
\multicolumn{1}{c}{$\beta_6$}              & 0 (-0.02,0.02)     & 0 (-0.09,0.07)      & -                  & -                 & 0.01 (-0.03,0.05)  & 0 (-0.39,0.42)  & 0.26 (-0.91,1.43)  \\
\multicolumn{1}{c}{$\beta_7$}              & 0 (-0.01,0.01)     & 0 (-0.06,0.06)      & -                  & -                 & 0.01 (-0.02,0.03)  & -0.02 (-0.40,0.33)  & -0.09 (-0.94,0.77) \\
\hline
\multicolumn{1}{c}{$\phi_1$}               & 0.53 (0.36,0.70)   & 0.08 (-0.13,0.29)   & 0.52 (0.34,0.67)   & 0.31 (0.13,0.49)   & -                  & -                  & -                  \\
\multicolumn{1}{c}{$\phi_2$}               & -                  & 0.18 (-0.02,0.39)   & -                  & 0.13 (-0.06,0.32)  & -                  & -                  & -                  \\
\multicolumn{1}{c}{$\phi_3$}               & -                  & 0.07 (-0.11,0.25)   & -                  & -0.05 (-0.25,0.15) & -                  & -                  & -                  \\
\multicolumn{1}{c}{$\phi_4$}               & -                  & 0.22 (0.04,0.39)    & -                  & 0.20 (0.01,0.40)   & -                  & -                  & -                  \\
\multicolumn{1}{c}{$\phi_5$}               & -                  & 0.09 (-0.08,0.26)   & -                  & 0.09 (-0.14, 0.32) & -                  & -                  & -                  \\
\multicolumn{1}{c}{$\phi_6$}               & -                  & 0.06 (-0.10,0.23)   & -                  & 0.07 (-0.15, 0.29) & -                  & -                  & -                  \\
\multicolumn{1}{c}{$\phi_7$}               & -                  & 0.06 (-0.10,0.22)   & -                  & 0.06 (-0.17,0.28)  & -                  & -                  & -                  \\
\hline
\multicolumn{1}{c}{p-value}                & 0.18               & 0.72                & \textless{}0.001   & 0.97              & \textless{}0.001   & \textless{}0.001   & \textless{}0.001  \\
\hline \hline
                                            & \multicolumn{7}{c}{Subject ID: 7708}                                                                                                                        \\ \cline{2-8} 
                                            & BDLM-AR(1)          & BDLM-AR(7)           & RegAR(1)             & RegAR(7)              & BDLagM              & BR-DLM              & NB-DLM              \\ \hline
\multicolumn{1}{c}{$\mu$}                  & 2.62 (1.73,3.50)   & 2.80 (-1.3,7.57)   & 2.76 (1.91,3.62)    & 4.64(2.38,6.89)     & 2.82 (2.25,3.39)    & 2.74 (2.17,3.30)    & 2.85 (2.25,3.44)   \\
\multicolumn{1}{c}{Total effect}           & -0.99 (-2.28,0.17) & -1.13 (-3.14,0.28) & -1.42 (-2.64,-0.19) & -1.72 (-3.14,-0.31) & -1.53 (-2.44,-0.61) & -1.38 (-2.32,-0.43) & -1.62 (-2.63,-0.6) \\
\multicolumn{1}{c}{Total carryover effect} & -0.36 (-1.58,0.50) & -0.40 (-1.96,0.52) & -                   & -                   & -0.45 (-2.59,1.66)  & -0.86 (-2.09,0.41)  & -0.46 (-2.73,1.80) \\
\multicolumn{1}{c}{$\beta_0$}              & -0.63 (-1.80,0.48) & -0.72 (-2.26,0.45) & -1.42 (-2.64,-0.19) & -1.72 (-3.14,-0.31) & -1.07 (-3.15,1.02)  & -0.52 (-1.73,0.45)  & -1.15 (-3.42,1.11) \\
\multicolumn{1}{c}{$\beta_1$}              & -0.22 (-1.03,0.45) & -0.32 (-1.46,0.39) & -                   & -                   & -0.02 (-2.46,2.43)  & -0.26 (-1.49,0.92)  & -0.11 (-3.22,2.98) \\
\multicolumn{1}{c}{$\beta_2$}              & -0.08 (-0.61,0.36) & -0.08 (-0.74,0.44) & -                   & -                   & -0.15 (-1.21,0.91)  & -0.13 (-1.33,1.14)  & 0.10 (-3.03,3.22)  \\
\multicolumn{1}{c}{$\beta_3$}              & -0.03 (-0.38,0.26) & -0.04 (-0.51,0.34) & -                   & -                   & -0.12 (-0.63,0.39)  & -0.08 (-1.27,1.18)  & 0.34 (-2.76,3.44)  \\
\multicolumn{1}{c}{$\beta_4$}              & -0.03 (-0.29,0.15) & -0.04 (-0.41,0.22) & -                   & -                   & -0.07 (-0.37,0.22)  & -0.29 (-1.63,0.86)  & -1.33 (-4.43,1.77) \\
\multicolumn{1}{c}{$\beta_5$}              & 0 (-0.15,0.13)     & 0.03 (-0.14,0.34)  & -                   & -                   & -0.04 (-0.22,0.13)  & 0.09 (-1.06,1.44)   & 0.99 (-2.14,4.11)  \\
\multicolumn{1}{c}{$\beta_6$}              & 0 (-0.10,0.09)     & 0.02 (-0.12,0.22)  & -                   & -                   & -0.03 (-0.13,0.08)  & -0.01 (-1.17,1.23)  & 0.13 (-2.95,3.22)  \\
\multicolumn{1}{c}{$\beta_7$}              & 0 (-0.06,0.05)     & 0.01 (-0.08,0.13)  & -                   & -                   & -0.02 (-0.08,0.05)  & -0.18 (-1.31,0.84)  & -0.58 (-2.85,1.69) \\
\hline
\multicolumn{1}{c}{$\phi_1$}               & 0.44 (0.26,0.62)   & 0.43 (0.22,0.64)   & 0.43 (0.24,0.59)    & 0.36 (0.17,0.54)     & -                   & -                   & -                  \\
\multicolumn{1}{c}{$\phi_2$}               & -                  & 0 (-0.22,0.23)     & -                   & -0.01 (-0.21,0.19)   & -                   & -                   & -                  \\
\multicolumn{1}{c}{$\phi_3$}               & -                  & -0.02 (-0.25,0.22) & -                   & 0 (-0.20,0.21)       & -                   & -                   & -                  \\
\multicolumn{1}{c}{$\phi_4$}               & -                  & 0.19 (-0.04,0.42)  & -                   & 0.13 (-0.07,0.33)    & -                   & -                   & -                  \\
\multicolumn{1}{c}{$\phi_5$}               & -                  & 0.09 (-0.15,0.32)  & -                   & 0.11 (-0.10,0.31)    & -                   & -                   & -                  \\
\multicolumn{1}{c}{$\phi_6$}               & -                  & 0.06 (-0.18,0.30)  & -                   & 0.06 (-0.16,0.28)    & -                   & -                   & -                  \\
\multicolumn{1}{c}{$\phi_7$}               & -                  & -0.05 (-0.28,0.17) & -                   & -0.09 (-0.29,0.12)   & -                   & -                   & -                  \\
\hline
\multicolumn{1}{c}{p-value}                & 0.76               & 0.83               & \textless{}0.001    & 0.91                & \textless{}0.001    & \textless{}0.001    & \textless{}0.001   \\
\hline
\end{tabular}}
\end{table}

\section{Discussion}
\label{s:discuss}

In this paper, we introduce a novel method to analyze data from N-of-1 trials. The method handles temporal correlation between measurements and carryover effects via distributed lag structure and parameters are estimated using Bayesian approach with (fused) ridge type regularization. From the design perspective, N-of-1 trial can be viewed as a multi-period crossover trial in an individual. Traditional crossover trial requires physical washout period to eliminate carryover effects, resulting in pauses of study intervention and potentially lengthening study duration. Instead of using physical washout period, our proposed method provides an alternative to address the carryover effects analytically, which can be applied to N-of-1 trial even without washout period. This is specifically suited to applications where outcomes are measured continuously over the study period.
Our simulation studies show that the proposed BDLM-AR model generally outperforms Koyck DLM, BDLagM, BR-DLM and NB-DLM in estimating  carryover effects while comparable in estimating the total effects.
Furthermore, we showed that BDLM-AR can simultaneously estimate autoregressive error. The advantage of BDLM-AR model increases when strong serial correlation exists.

We adopt a Bayesian estimation framework, which facilitates modeling and inference of N-of-1 trial data in several ways. First, a key in modeling carryover effects in N-of-1 trial data is to address multicollinearity in the lag treatment effects. Under a Bayesian framework, we achieve this by postulating a prior precision matrix on the lag coefficients to provide the appropriate constraints on the lag coefficients. Specifically, the designed form of this prior precision matrix is motivated by and connected to a fused ridge penalized estimation procedure (\ref{eq:penalty}), which imposes shrinkage and smoothness of the lag coefficients; see Appendix A for details.
Second, while cross validation is often a method of choice in tuning the penalty terms ($\lambda_i$ and $\lambda_i^*$ via $\gamma_1$ and $\gamma_2$) in penalized estimation, it is not feasible to split the sample at random time points because of the temporal order in N-of-1 trial data. Bayesian formulation provides a natural way to tune the penalties in a data-driven manner via posterior inference.
Third, we have applied our model to a real application of using white light therapy for depressive symptoms, along with other Bayesian approaches (BDLagM, BR-DLM, NB-DLM). These approaches allow for using posterior credible intervals of individual lag coefficients as inferential tools. While there are varying degrees of variability, the proposed BDLM-AR gives the shortest intervals. The posterior intervals offer additional insights on the whole time course of treatment effect.

\section*{Data availability}
R code for our Bayesian distributed lag model and corresponding simulations is available via the following GitHub repository,
https://github.com/williammomo/BDLM-AR.

\section*{Appendix}
\subsection*{Appendix A}\label{AppendixA}
Let $\pi(A|\cdot)$ denote conditional distribution of $A$ given all other variables in the model. The posterior distribution of $\tilde{\boldsymbol{\beta}}=(\mu, \boldsymbol\beta')'$ is
\begin{equation*}
\begin{aligned}
\pi(\tilde{\boldsymbol{\beta}} \mid \cdot) & \propto \pi(\boldsymbol{Y}|\tilde{\boldsymbol{X}},\tilde{\boldsymbol{\beta}},\sigma^2,\boldsymbol{\phi},\boldsymbol{\gamma})\pi(\tilde{\boldsymbol{\beta}}|\sigma^2,\boldsymbol{\gamma}) \\
&\propto \exp \left[-\frac{1}{2 \sigma^{2}}(\boldsymbol{Y}^*-\tilde{\boldsymbol{X}}^* \tilde{\boldsymbol{\beta}})'(\boldsymbol{Y}^*-\tilde{\boldsymbol{X}}^* \tilde{\boldsymbol{\beta}})\right] \exp \left[-\frac{1}{2 \sigma^{2}}\tilde{\boldsymbol{\beta}}'\tilde{\boldsymbol{\Omega}}(\boldsymbol{\gamma})\tilde{\boldsymbol{\beta}}\right] \\
&\propto \exp \left[-\frac{1}{2 \sigma^{2}}(\boldsymbol{Y}^*-\tilde{\boldsymbol{X}}^* \boldsymbol{\beta})'(\boldsymbol{Y}^*-\tilde{\boldsymbol{X}}^* \boldsymbol{\beta}) - \frac{1}{2\sigma^2}c_0\mu^2 - \frac{1}{2 \sigma^{2}}\sum_{l=0}^L \lambda_l\beta_l^2 - \frac{1}{2\sigma^2} \sum_{l=0}^{L}\lambda^*_l(\beta_l-\beta_{l+1})^2\right].
\end{aligned}
\end{equation*}
The MAP estimate of $\tilde{\boldsymbol{\beta}}$ is 
\begin{equation*}
\begin{aligned} 
\hat{\boldsymbol\beta}_{\text{MAP}} &= 
\arg\max_{\tilde{\boldsymbol{\beta}}}\pi(\tilde{\boldsymbol{\beta}} \mid \cdot)\\
&= \arg\min_{\tilde{\boldsymbol{\beta}}}(\boldsymbol{Y}^*-\tilde{\boldsymbol{X}}^*\tilde{\boldsymbol{\beta}})' (\boldsymbol{Y}^* - \tilde{\boldsymbol{X}}^* \tilde{\boldsymbol{\beta}}) + c_0 \mu^2 + \sum_{l=0}^L \lambda_l\beta_l^2 + \sum_{l=0}^{L}\lambda^*_l(\beta_l-\beta_{l+1})^2
\end{aligned}
\end{equation*}
Note that the model variance $\sigma^2$ is included in the prior distribution of $\tilde{\boldsymbol\beta}$ as a scaling parameter. In this way, the ridge and smoothness penalties are only controlled by $\lambda_l$ and $\lambda^*_l$, without depending on the model variance and prior covariance matrix of $\tilde{\boldsymbol{\beta}}$. 

\subsection*{Appendix B}\label{AppendixB}
\begin{algorithm}[H]
\SetAlgoLined
 \textbf{Step 1.} Set initial values for $\tilde{\boldsymbol{\beta}}$, $\sigma^2$, $\boldsymbol{\phi}$ and $\boldsymbol{\gamma}$\;
 \For{$i \gets 1$ to $n_{\text{iteration}}$}{
  \textbf{Step 2.} Given current value of $\boldsymbol{\phi}$, transform $\boldsymbol{Y}$, $\tilde{\boldsymbol{X}}$ to $\boldsymbol{Y}^*$, $\tilde{\boldsymbol{X}}^*$ as described in equation (3) of Section 2.1; Also construct precision matrix $\tilde{\boldsymbol{\Omega}}(\boldsymbol{\gamma})$ based on $\boldsymbol{\gamma}$ as described in Section 2.2\;
  \textbf{Step 3.} Conditional on current values of $\boldsymbol{Y}^*$, $\tilde{\boldsymbol{X}}^*$, $\sigma^2$ and $\tilde{\boldsymbol{\Omega}}(\boldsymbol{\gamma})$, update $\tilde{\boldsymbol{\beta}}$ based on $\pi(\tilde{\boldsymbol{\beta}} \mid \boldsymbol{Y}^*,\tilde{\boldsymbol{X}}^*,\sigma^2,\boldsymbol{\phi},\boldsymbol{\gamma})$\;
  \textbf{Step 4.} Conditional on current values of $\boldsymbol{Y}^*$, $\tilde{\boldsymbol{X}}^*$, $\tilde{\boldsymbol{\beta}}$, $\boldsymbol{\phi}$ and $\tilde{\boldsymbol{\Omega}}(\boldsymbol{\gamma})$, update $\sigma^2$ based on $\pi(\sigma^2 \mid \boldsymbol{Y}^*,\tilde{\boldsymbol{X}}^*,\tilde{\boldsymbol{\beta}},\boldsymbol{\phi},\boldsymbol{\gamma})$\;
  \textbf{Step 5.} Update $\boldsymbol{\epsilon}^*$ conditional on current value of $\tilde{\boldsymbol{\beta}}$ and $\boldsymbol{Y}$, $\tilde{\boldsymbol{X}}$. Then update $\boldsymbol{\phi}$ based on $\pi(\boldsymbol{\phi} \mid \boldsymbol{Y},\tilde{\boldsymbol{X}},\tilde{\boldsymbol{\beta}},\sigma^2,\boldsymbol{\gamma})$. Reject samples if the roots of $\boldsymbol{\phi}(L)$ lie outside the unit circle\;
  \textbf{Step 6.} Update $(\gamma_1,\gamma_2)$ based on $\pi(\boldsymbol{\gamma} \mid \tilde{\boldsymbol{\beta}}, \sigma^2)$. Sample a proposal $\gamma_i^*$ by $\gamma_i^* = {\gamma_i}+a*U(-1,1)$ for $i =1,2$. $a$ is an adjustable step size. Compute the ratio
  \begin{equation*}
  R_{\gamma}=\frac{\pi(\boldsymbol{\gamma}^* \mid \tilde{\boldsymbol{\beta}}, \sigma^2)}{\pi(\boldsymbol{\gamma} \mid \tilde{\boldsymbol{\beta}}, \sigma^2)} 
  \end{equation*}
  \If {$\tilde{\boldsymbol{\Omega}}(\boldsymbol{\gamma}^*)$ is positive definite}{
  update $\boldsymbol{\gamma} = \boldsymbol{\gamma}^*$ with probability min(1, $R_{\gamma}$)\;
  }

 }
 \caption{The hybrid Metropolis-Hastings/Gibbs algorithm}
\end{algorithm}

\clearpage

\nocite{*}% Show all bib entries - both cited and uncited; comment this line to view only cited bib entries;
\bibliography{references}%
\bibliographystyle{ieeetr}

\clearpage

\end{document}